\documentclass[a4paper,11pt]{article}
\pdfoutput=1 

\usepackage{jcappub} 

\usepackage[T1]{fontenc} 
\usepackage{graphicx}
\usepackage{natbib}
\usepackage{txfonts}
\usepackage{enumitem}
\usepackage[lofdepth,lotdepth]{subfig}

\title{\boldmath Extragalactic neutrinos as tracers of Dark Matter?}

\author[a,1]{A.V. Penacchioni \note{Corresponding author.}}
\author[a,b]{O. Civitarese}

\affiliation[a]{IFLP (CONICET), La Plata, Argentina.}
\affiliation[b]{Department of Physics, University of La Plata (UNLP), 49 y 115 cc. 67, 1900 La Plata, Argentina.}

\emailAdd{ana.penachioni@fisica.unlp.edu.ar}
\emailAdd{osvaldo.civitarese@fisica.unlp.edu.ar}

\abstract{Neutrinos produced in extragalactic sources may experience flavor-oscillations and decoherence on their way to Earth due to their interaction with dark matter (DM). As a result, they may be detected in pointer-states other than the flavor states at the source. The oscillation pattern and the structure of the pointer-states can give us information on the characteristics of the DM and the kind of interaction that has taken place. From this perspective, neutrinos can be viewed as DM-tracers.\\
We study the local evolution of neutrino flavor-eigenstates due to local effects produced by the presence of DM. To explore the sensitivity of the model, we consider different DM density profiles, masses and interactions. \\
Starting from the eigenstates of the neutrino-mass Hamiltonian, we construct the flavor-states with the neutrino mixing-matrix in vacuum. We then include local interactions with DM, acting along the neutrino path towards the Earth, and analyse the resulting probabilities. In doing so, we adopt different DM density profiles, e.g. a constant, a local isotropic and a Navarro-Frenk-White density distribution. Finally, by following the time evolution of the flavor-states, we identify pointer-states and interpret the results in terms of the adopted DM model.\\
Due to the interaction with DM, neutrinos experience the MSW effect, the extent of which depends on the DM density profile. The interaction with DM produces the enhancement or suppression of oscillations. Decoherence effects may take place.  \\
We model the time evolution of extragalactic neutrino flavor-states by letting them interact with DM. The features of the calculated response seem to support the notion that these neutrinos can be taken as DM tracers. From a theoretical point of view, the coexistence and/or competition of decoherence and MSW effects is sustained by the results. }

\keywords{neutrino theory, neutrino properties, dark matter theory }

\begin{document}
\maketitle
\flushbottom

\section{Introduction}
Dark matter (DM) \citep{2007PhRvD..76c3007M,2009PhRvD..79d3508D} accounts for the largest fraction of constituents of the Universe. The characterization of the properties of DM and dark energy, the composition of which is still unknown, is one of the hot topics in today's Physics, with direct impact on astrophysics. Although it has not yet been detected directly \citep{2009NJPh...11j5011S,2008PhRvD..78a5020H}, we may infer its existence from the shape, motion and distribution of baryonic matter in the Universe. For example, the amount of matter we see in a galaxy-cluster is not enough to keep the galaxies together by gravity alone, so there must be some additional matter present.

Despite the ignorance about the actual composition of DM, many candidates have been proposed in the last decades, each of them interacting in its own particular way. 
The most widely accepted candidates are leptons, known as Weakly Interacting Massive Particles (WIMPs) \citep{2008PhRvL.101i1301A,2009PhRvL.102a1301A}. These would be completely different from normal matter and could interact via gravity and the weak force. If DM is composed by WIMPs, then these particles should be much more abundant than normal matter in order to describe what we observe in the Universe. We should be able to detect them through their collisions with particles in detectors.
Other DM candidates are axions. They are low-mass, slow-moving neutral particles that interact via the weak force, which may condensate and decay into a pair of photons \citep{2012PhRvD..85f3520E}.
There are many other candidates like the MAssive Compact Halo Objects (MACHOs) \citep{2018PhRvD..98l3523J}, that could be Neutron Stars (NS), White and Brown Dwarfs and other compact objects made of ordinary matter that produce very little light (or no light at all). Other DM candidates are the Supersymmetric (SUSY) particles \citep{1996PhR...267..195J}, the Kaluza-Klein particle, etc. 

Neutrinos are produced in almost all the astrophysical sources and events, both galactic and extragalactic. Examples of these are solar flares, pulsars, supernova remnants, mass accretion onto a black hole (BH) \citep{2019MNRAS.485.3352L}, gamma-ray bursts (GRBs), microquasars, X-ray binaries and blazars \citep{2007JPhCS..60....8D}. As a consequence of their extremely weak interaction with other particles, neutrinos are very good carriers of information about the processes that originated them.

Since they have non-zero mass, neutrinos experience oscillations between flavor states ($\nu_e$, $\nu_{\mu}$ and $\nu_{\tau}$). Even if we cannot determine the value of the neutrino mass, because the oscillation experiments measure squared-mass differences between mass eigenstates, it is customary to introduce the concept of mass hierarchy \citep{2006NuPhB.734...24P}. Masses are usually ordered according to hierarchies as: Normal (NH) $\rightarrow$ $m_1 \lesssim m_2 << m_3$, Inverted (IH) $\rightarrow$ $m_1 << m_2 \lesssim m_3$ and Degenerate (DH) $\rightarrow$ $m_1 \approx m_2 \approx m_3$ hierarchies.

The interaction of neutrinos with electrons generates significant changes in the masses $m_i$, as well as in the composition of the flavor states. This is known as the Mikheyev-Smirnov-Wolfenstein (MSW) effect \citep{2005PhST..121...57S,2016EPJC...76..339K}.

When neutrinos are produced in very distant (extragalactic) sources, and due to their interaction with the intergalactic medium, either with other particles or fields, they may experience decoherence \citep{2007dqct.book.....S}. As a result, neutrinos may be detected in "pointer"-states \citep{2019ApJ...872...73P}. This effect may play a role in the analysis of neutrino signals in experiments like Ice Cube \citep{2018arXiv181107979I}, ANTARES \citep{2017PhLB..769..249A}, KM3NeT \citep{2018arXiv181008499T}, etc. 

In this work we focus on the interaction of neutrinos produced in astrophysical sources with DM and study the possibility of using them as DM-tracers. We develop a scheme based on the evolution of flavor-eigenstates due to local effects associated to the presence of DM. The procedure consists of the following steps: 

\begin{enumerate}[label=\roman*)]
    \item construction of the flavor states from the neutrino-mixing matrix in vacuum, starting from the eigenstates of the mass Hamiltonian $H_m$;
    \item inclusion of local interactions with DM and subsequent analysis of the resulting MSW effect;
    \item follow-up of the time-evolution of the neutrino flavor-states and identification of the pointer states.
\end{enumerate}

In dealing with the MSW we follow the work of \cite{2016PhRvD..94l3001D} and take DM: i) isotropically distributed in the Universe, and ii) localized in the halo of the galaxy, having a Navarro-Frenk-White (NFW) density profile \citep{1996ApJ...462..563N,1997ApJ...490..493N}. Furthermore, we take also a constant DM density distribution. Concerning the time evolution we introduce a set of intermediate points along the neutrino path from the source to the detector. 

The paper is organised as follows: in Section \ref{sec:oscillations} we introduce the model and describe neutrino oscillations in vacuum (Subsection \ref{2.1}) and in presence of DM (Subsection \ref{2.2}). Results and discussions are presented in Section \ref{sec:discussion}. Finally, we draw our conclusions in Section \ref{sec:conclusions}.

\section{Neutrino oscillations}\label{sec:oscillations}

Neutrinos experience flavor oscillations. Each of the neutrino flavor states ($\nu_e$, $\nu_{\mu}$, $\nu_{\tau}$) is a superposition of neutrino mass-eigenstates ($\mid m_i \rangle$, $i=1,2,3$). As neutrinos of flavor $\alpha$ propagate through space their composition changes due to the non-vanishing mass differences between the mass eigenvalues $m_i$.

Neutrinos of flavor $\alpha$ are written as linear combinations of the mass-eigenstates as:
\begin{equation}
\mid \nu_{\alpha} \rangle= \sum_{i} U_{\alpha \, i} \mid m_i \rangle,
\end{equation}
where $U$ is the Pontecorvo-Maki-Nakagawa-Sakata matrix (PMNS matrix) \citep{2000hep.ph....1311B}.  

The Hamiltonian in the mass basis, $H_m$, has a diagonal form
\begin{equation}
H_m = 
\begin{bmatrix}
m_1 & 0 & 0\\
0 & m_2 & 0\\
0 & 0 & m_3\\
\end{bmatrix}
.
\end{equation}

The Hamiltonian in the flavor basis has a non-diagonal form and is written as
\begin{equation}\label{Hf}
H_f = U \, H_m \, U^{\dagger},
\end{equation}
where the elements of the mixing-matrix U are given by
\begin{align}\label{eq:U}
U=
\begin{bmatrix}
c_{12} c_{13} & s_{12} c_{13} & s_{13} \, e^{-i\delta} \\
  -s_{12} c_{23}-c_{12} s_{23} s_{13} \, e^{i\delta} & c_{12} c_{23}-s_{12} s_{23} s_{13} \, e^{i\delta} & s_{23} c_{13}  \\
  s_{12} s_{23}-c_{12} c_{23} s_{13} \, e^{i\delta} & -c_{12} s_{23}-s_{12} c_{23} s_{13} \, e^{i\delta} & c_{23} c_{13}  \\
\end{bmatrix}
\times \rm{diag}\left(1, \, exp(i\frac{\alpha_{21}}{2}), \,exp(i\frac{\alpha_{31}}{2}) \right).
\end{align}
The quantities $c_{ij}$ ($s_{ij}$) are the cosine (sine) of the mixing angles $\theta_{ij}$. The Dirac ($\delta$) and Majorana ($\alpha$) phases are set to 0.

The interaction of neutrinos with DM is represented by the effective potential
\begin{equation}\label{V}
V= \lambda \, G_F \, \frac{\rho(r)}{m_{DM}} \, \Lambda,
\end{equation}
where $\lambda$ is a dimensionless scale parameter, $G_F=8.963\times 10^{-44}$ MeV cm$^3$ is the Fermi constant, $\rho(r)$ is the DM density distribution, $m_{DM}$ is the DM mass in units of energy, and $\Lambda$ is a $3 \times 3$ matrix. 

\begin{figure}
\centering
\includegraphics[width=0.6\linewidth]{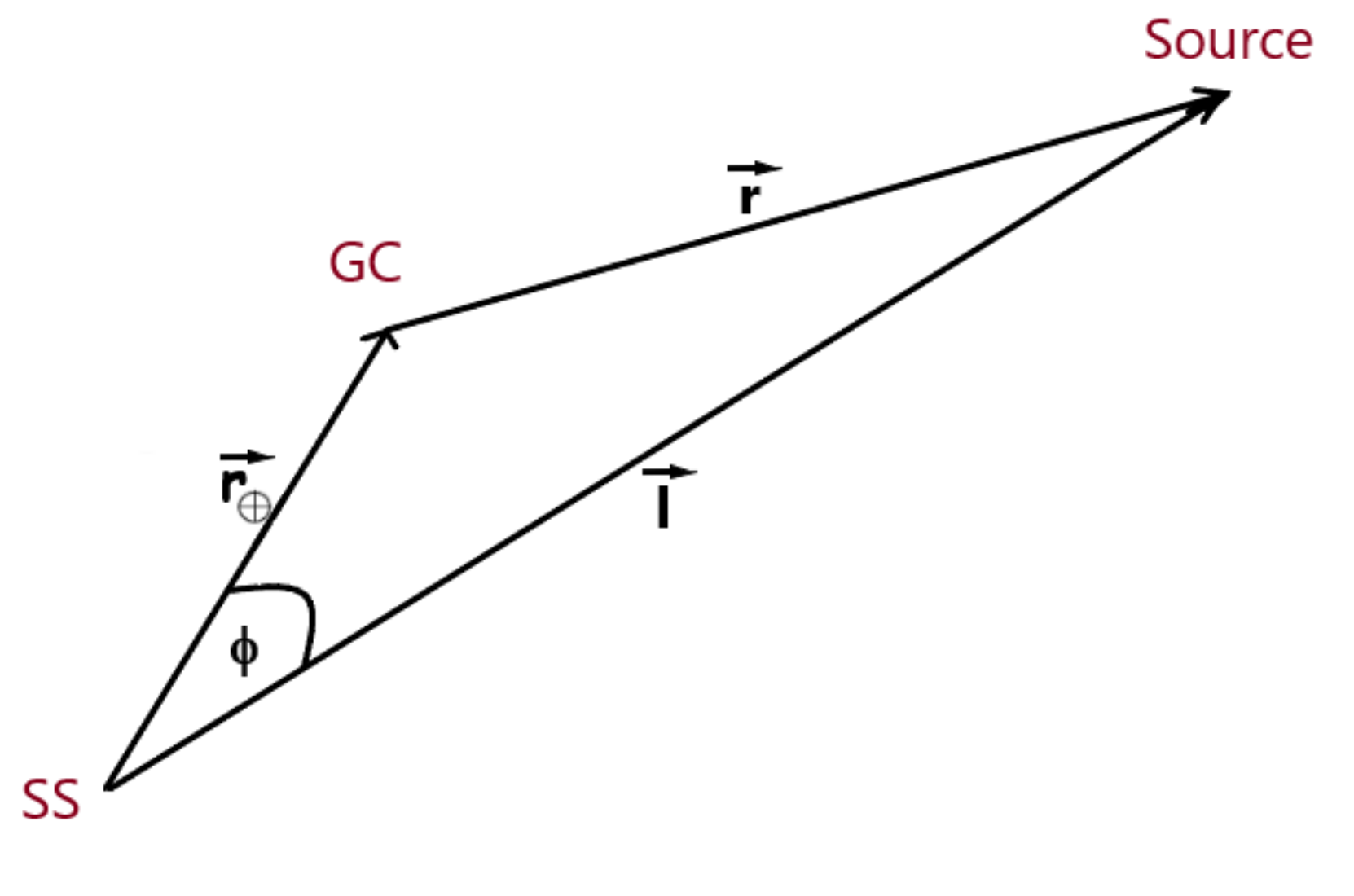}
\caption{ Scheme showing the distances considered in our calculations. SS stands for Solar System and GC stands for Galactic Center.}
\label{fig:diagrama}
\end{figure}

\begin{figure}
\centering
\includegraphics[width=0.6\linewidth]{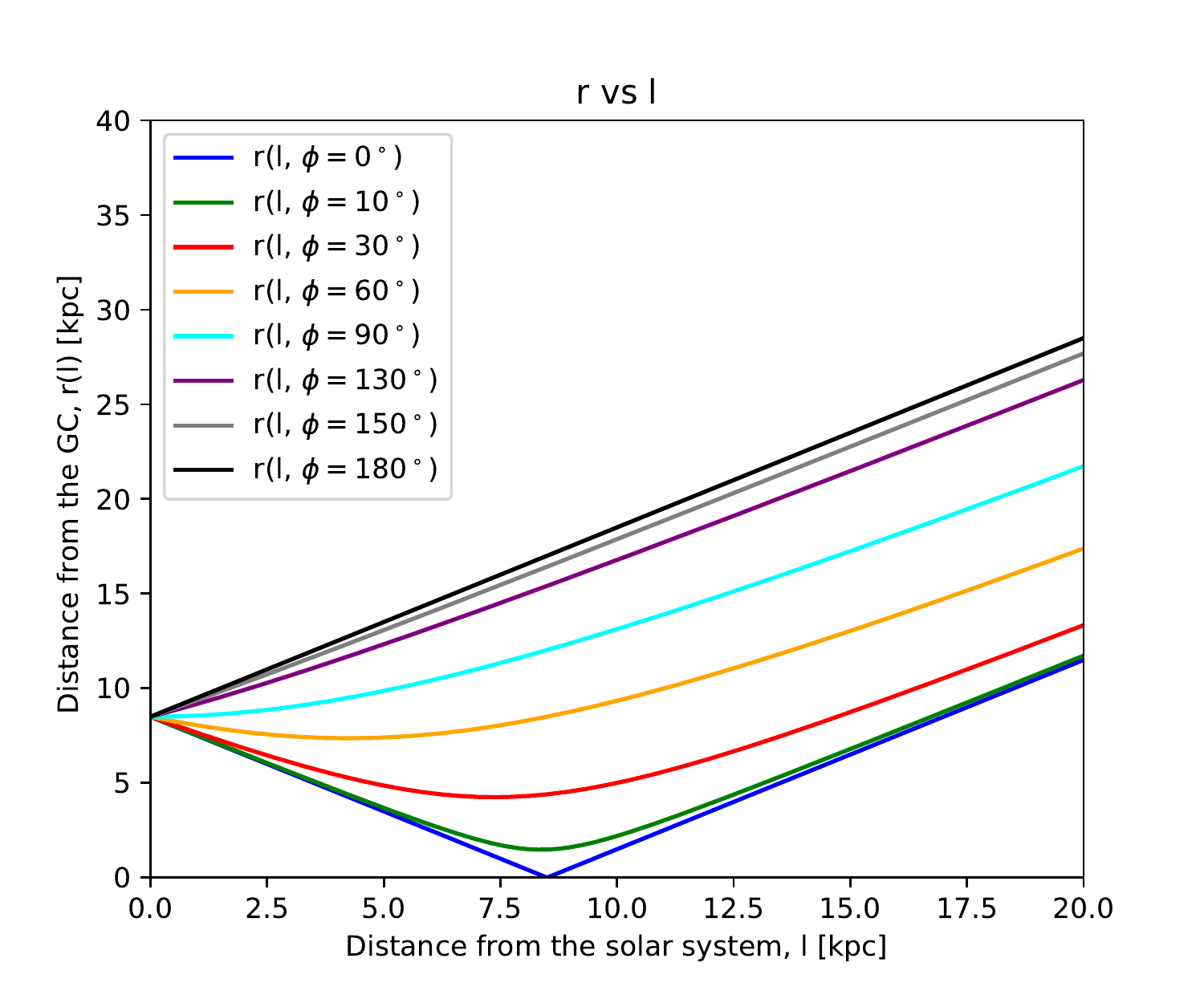}
\caption{Distance $|r|$ from the GC to the source as a function of $l$, for different values of the angle $\phi$ (See Eq.\ref{eq:r}), from $\phi=0^\circ$ (bottom) to $\phi=180^\circ$ (top).}
\label{fig:rversusl}
\end{figure}

\begin{figure}
\centering
\includegraphics[width=0.6\linewidth]{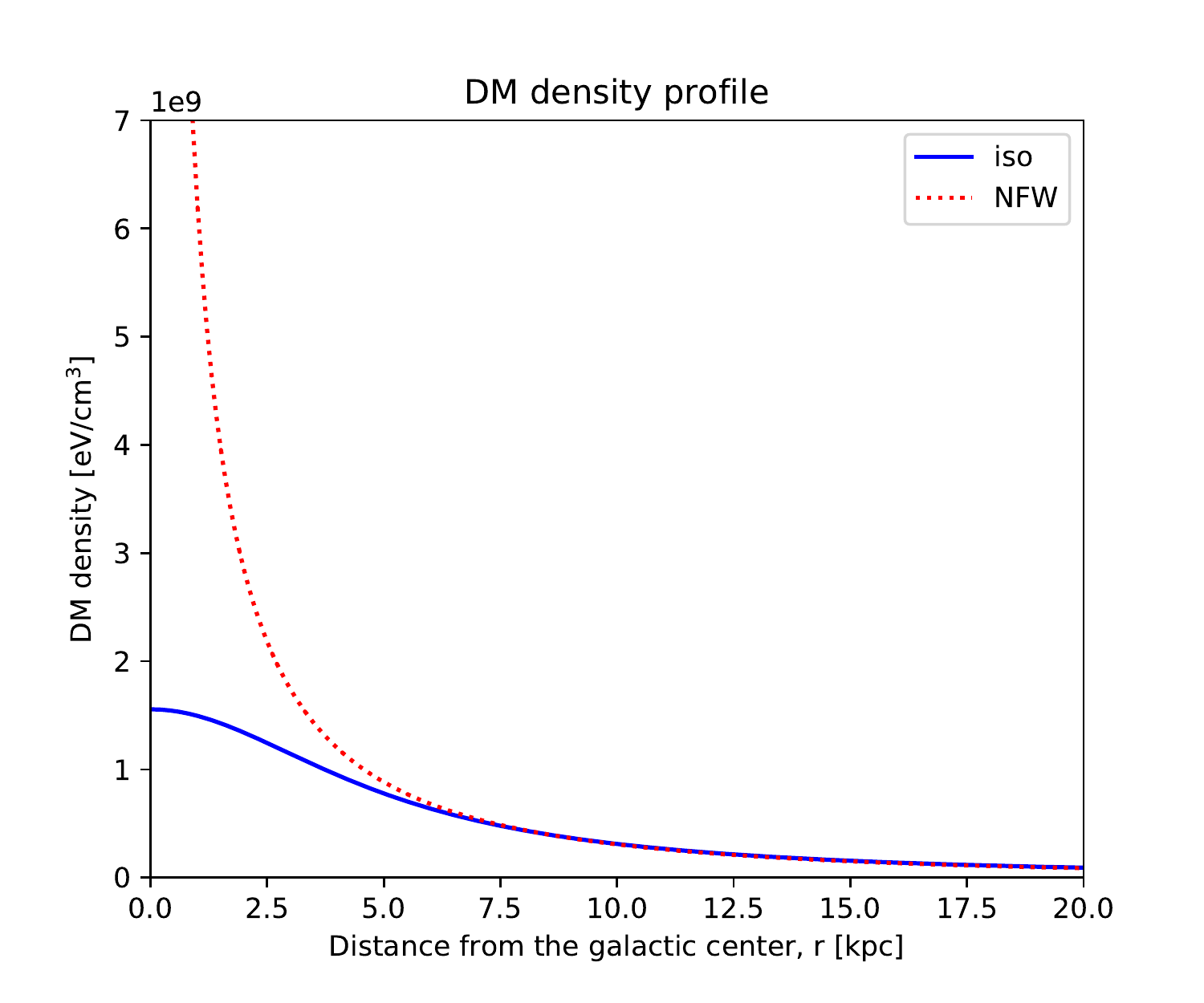}
\caption{DM density profiles considered in our calculations as a function of $r$ for $\phi=0$: isotropic (solid line, see Eq. \ref{eq:iso}) and Navarro-Frenk-White (dotted line, see Eq. \ref{eq:NFW}).}
\label{fig:rhoDM}
\end{figure}

We use three DM-distributions in our treatment. One is the isotropic distribution of relic neutrinos homogeneously distributed in the Universe
\begin{equation}\label{eq:iso}
\rho_{\rm{iso}}(r)=\rho_{\oplus}\left(\frac{1+(r_{\oplus}/r_s)^2}{1+(r/r_s)^2}\right)
\end{equation}
with $r_s=5 \, \rm{kpc}$. $\rho_{\oplus}=0.4$ GeV cm$^{-3}$ is the local DM density and $r_{\oplus}=8.5$ kpc is the distance of the Solar System to the Galactic Center (GC). The second one is the NFW distribution, for DM neutrinos in the halo of our galaxy
\begin{equation}\label{eq:NFW}
\rho_{\rm{NFW}}(r)=\rho_{\oplus}\left(\frac{r_{\oplus}}{r}\right)\left(\frac{1+(r_{\oplus}/r_s)}{1+(r/r_s)}\right)^2,
\end{equation}
with  $r_s=20 \, \rm{kpc}$, and the third one is a constant density distribution, $\rho_{const}(r)=\rho_{\oplus}$.
If $l$ is the distance from the solar system to the source and $\phi$ is the angle between $l$ and $r_{\oplus}$, then
\begin{equation}\label{eq:r}
|r|=\sqrt{r_{\oplus}^2+l^2-2\,l\,r_{\oplus} \, \rm{cos}\,\phi}
\end{equation}
is the distance from the GC to the source, as shown in the scheme depicted in Figure \ref{fig:diagrama}. Actual values of $r$ as a function of $l$ and $\phi$ are shown in Figure \ref{fig:rversusl}. The DM density profiles are shown in Figure \ref{fig:rhoDM} as a function of $r$ for $\phi=0$. 

\subsection{Neutrino oscillations in vacuum}\label{2.1}

Neutrino flavor-states evolve with time according to 
\begin{equation}
\mid \nu_{\alpha}(t) \rangle = \sum_k U_{\alpha k}(t) e^{-iE_it/\hbar} \mid m_k \rangle,
\end{equation}
where $E_i=\sqrt{p^2 c^2 + m_i^2 c^4}=pc \left(1+\frac{m_i^2c^4}{p^2c^2}\right)^{1/2} \approx pc+\frac{m_i^2 c^4}{2E}$ is the channel energy and $E$ is the energy of the flavor state. The states $\mid \nu_{\alpha}(t) \rangle $ are eigenstates of the flavor Hamiltonian (Eq. \ref{Hf}). Defining $\delta^2_k(t) \equiv \frac{m_k^2c^4 t}{2E\hbar}$, we have 
\begin{equation}
    \mid \nu_{\alpha}(t) \rangle = e^{-ipct/\hbar} e^{-i\delta_1^2(t)} \sum_k U_{\alpha k} e^{-i(\delta_k^2(t)-\delta_1^2(t))} \mid m_k \rangle.
\end{equation}
In order to calculate the survival and transition probabilities we compute the amplitudes $A_{\alpha \beta}$, with $\alpha$, $\beta= e, \mu, \tau$. If we take the flavor $\alpha$ at $t=0$, the transition amplitude to the flavor $\beta$ at time $t$ is:

\begin{equation}
    A_{\alpha \beta} (t) =\langle \nu_{\beta} (t) \mid \nu_{\alpha} (0) \rangle =  e^{i\delta_1^2(t)}\sum_k U_{\alpha k} U_{\beta k}^* e^{i(\delta_k^2(t)-\delta_1^2(t))}.
\end{equation}
In this expression $U_{\alpha k}$ is the original mixing matrix, defined in Eq. \ref{eq:U}.
The probability is therefore given by
\begin{eqnarray}
P_{\alpha \beta}(t)&=& |A_{\alpha \beta} (t)|^2= (\rm{Re}\,A_{\alpha \beta}(t))^2+(\rm{Im}\,A_{\alpha \beta} (t))^2 \nonumber \\
&=&\sum_{k k'} U_{\alpha k}U^*_{\beta k} U^*_{\alpha k'} U_{\beta k'} \rm{cos}(\delta_k^2(t)-\delta_{k'}^2(t)), \nonumber \\
&&
\end{eqnarray}
and it measures the survival ($\alpha=\beta$) or disappearance ($\alpha \neq \beta$) of a given flavor. Obviously, $P_{\alpha \alpha}+ \sum_{\beta \neq \alpha} P_{\alpha \beta}=1$ for all $t$.

Neutrinos from the source are emitted at $t=0$ and travel towards the Earth at nearly the speed of light. If the source is located at a distance $L_{\rm{max}}$ from the solar system, then $l$ is related to $t$ by
\begin{equation}\label{eq:l}
l=L_{\rm{max}}-ct. 
\end{equation}

\subsection{Neutrino oscillations in a DM environment}\label{2.2}
We are interested in calculating the neutrino-oscillation probabilities as a function of time in presence of DM, that is considering the MSW effect \citep{2016EPJC...76..339K} taking the DM as a source. We consider for our calculations different values of the DM mass $m_{DM}$.
We can obtain the flavor-transition probabilities by diagonalizing the new Hamiltonian \citep{2016PhRvD..94l3001D}, 
\begin{equation}\label{H}
H= H_f+V(l,\phi).
\end{equation}
In this expression, $V(l,\phi)$ is the effective potential (see Eq. \ref{V}) that depends on the spatial coordinates through the DM distribution $\rho(r)$.
When using the standard approximation ($E>> m_ic^2$) we write for the mass Hamiltonian
\begin{equation}
H_m= \frac{1}{2 E_{\nu}}
\begin{bmatrix}
0 & 0 & 0\\
0 & \Delta m_{12}^2 & 0\\
0 & 0 & \Delta m_{13}^2\\
\end{bmatrix}
,
\end{equation}
where $\Delta m_{1j}^2 = (m_j^2-m_1^2)c^4$ are squared mass differences and $E_{\nu}$ is the neutrino energy.
We want to study the oscillation pattern of a neutrino from the moment it is emitted at a source located at a distance $L_{max}$ until it reaches the detector on Earth. Let us assume that the neutrino is created in the electron flavor at time $t=t_0$. The total Hamiltonian at the source is given by the flavor Hamiltonian in vacuum plus the effective potential at the source, that is $H(t_0)= H_f+V(L_{max},\phi)$. We need to diagonalize the Hamiltonian $H(t_0)$ in order to find the matrix of eigenvectors $W$ and the eigenvalues $w_i$. If now we transform back to the mass basis using 
\begin{equation}\label{H_m'}
    H_m'=W H(t_0) W^{-1},
\end{equation}
we obtain a new mass Hamiltonian that differs from the original one due to the presence of the potential $V$ in Eq. \ref{H}. This is due to the local dependence of the interaction.

As neutrinos move a distance $\Delta l$ towards the Earth (which is situated at $l=0$, see Figure \ref{fig:diagrama}), in a time $\Delta t= \Delta l /c$, the new Hamiltonian becomes $H(t_0+\Delta t)=H(t_0)+V(L_{max}-\Delta l,\phi)$. If we apply this procedure recursively we will obtain a Hamiltonian at Earth that will keep memory of the distribution of DM along the neutrino path. Furthermore, when we calculate the survival and disappearance probabilities for each flavor as a function of time we will get an oscillation pattern that is entirely dependent on the value of the effective potential at each point along the trajectory.

\section{Results and discussion}\label{sec:discussion}

\begin{figure}
\centering
\includegraphics[width=0.6\textwidth]{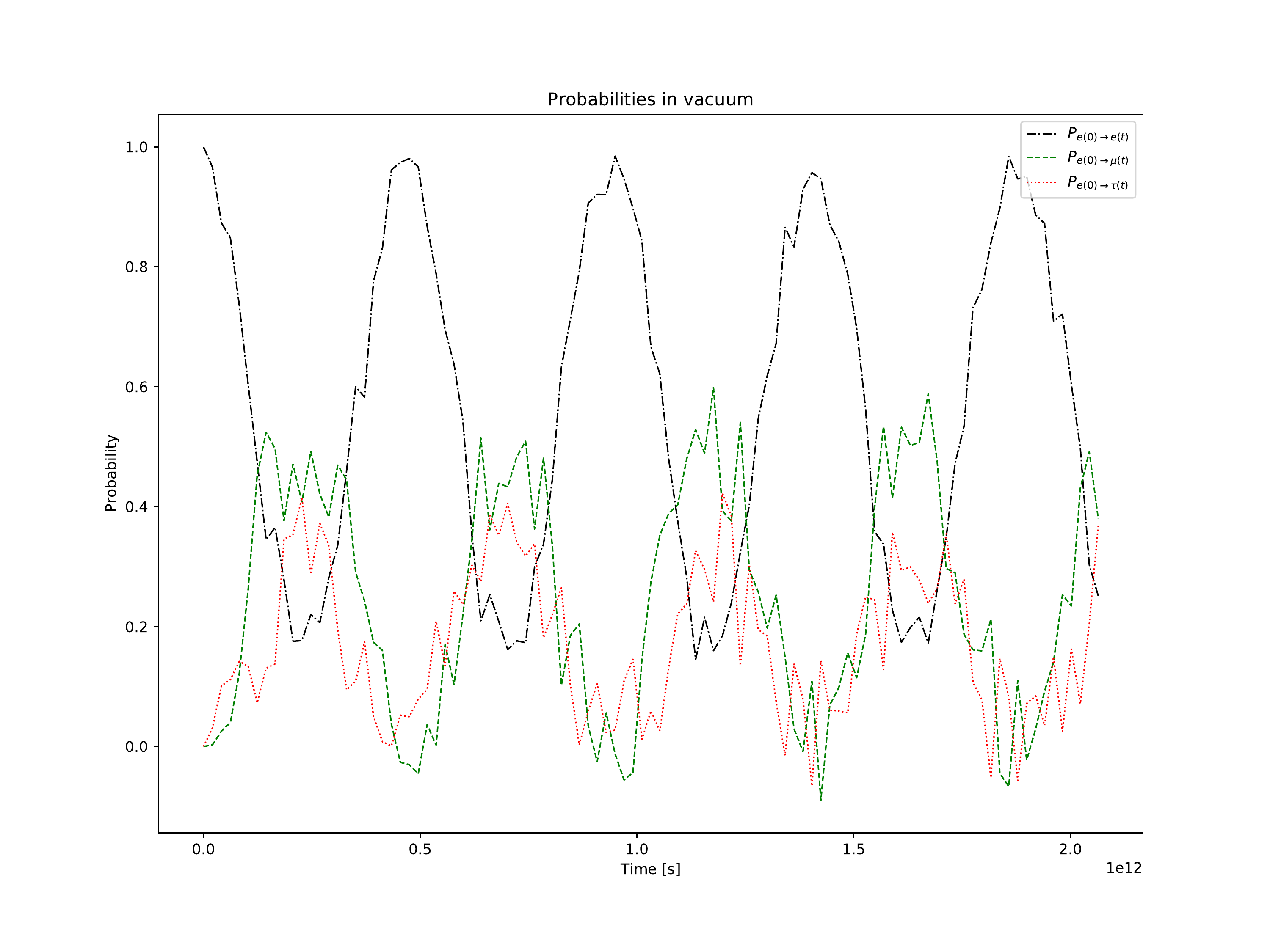}
\caption{Transition probabilities of finding the neutrino in the electron (black dash-dotted line), $\mu$ (green dashed line) or $\tau$ (red dotted line) flavor in vacuum as a function of time.}
\label{fig:Prob_vacio}
\end{figure}

Figure \ref{fig:Prob_vacio} shows the results of the calculations of the survival ($\nu_e$ $\rightarrow$ $\nu_e$)  and disappearance ($\nu_e \rightarrow \nu_{\mu (\tau)}$) probabilities in vacuum. The results have been obtained by taking $L_{\rm{max}}=20$ kpc, neglecting the distance between the Earth and the Sun ($\approx 4.86 \times 10^{-9}$ kpc) as compared with $r_{\oplus}$, setting $\phi=0$ (see Figure \ref{fig:rversusl}), taking the emission of $E_{\nu}=1$ MeV electron-neutrinos from the source at $t=0$ and adopting the NH for the initial mass-eigenstates.

From the expressions presented in Section \ref{sec:oscillations}, there are several unknown quantities to be fixed in order to extract, quantitatively, some information about DM when analysing neutrino's spectra. They are the mass of the DM particles, $m_{DM}$, the DM density profile $\rho(r)$, the texture of the matrix $\Lambda$, and the value of the dimensionless parameter $\lambda$ which renormalises the Fermi constant. The values of $m_{DM}$ can vary in a wide range, and DM may be uniformly or locally distributed in the Universe. The choice of the texture of the matrix $\Lambda$ is completely arbitrary, and to fix some representative values we have followed the analysis of \citep{2016PhRvD..94l3001D} as a guide. The parameter $\lambda$ is also unknown, so it will be varied in a large range, too.

As a first step, we fix the mass of the DM particles to the value $m_{DM}=10^{-3}$ eV, $\lambda=1$, and take the constant value $\rho_{\oplus}$ for the DM density distribution. The remaining degree of freedom is the texture of the matrix $\Lambda$, for which we take the following ansatzes: \\
$\Lambda_{(a,b,c)}= 
\begin{bmatrix}
1 & 0 & 0\\
0 & 1 & 0\\
0 & 0 & 1\\
\end{bmatrix}$,
$\begin{bmatrix}
0 & 0 & 1\\
0 & 0 & 0\\
1 & 0 & 0\\
\end{bmatrix}$, and
$\begin{bmatrix}
1 & 1 & 1\\
1 & 1 & 1\\
1 & 1 & 1\\
\end{bmatrix}$.

\begin{figure*}
\subfloat[][]{
\includegraphics[width=0.33\textwidth]{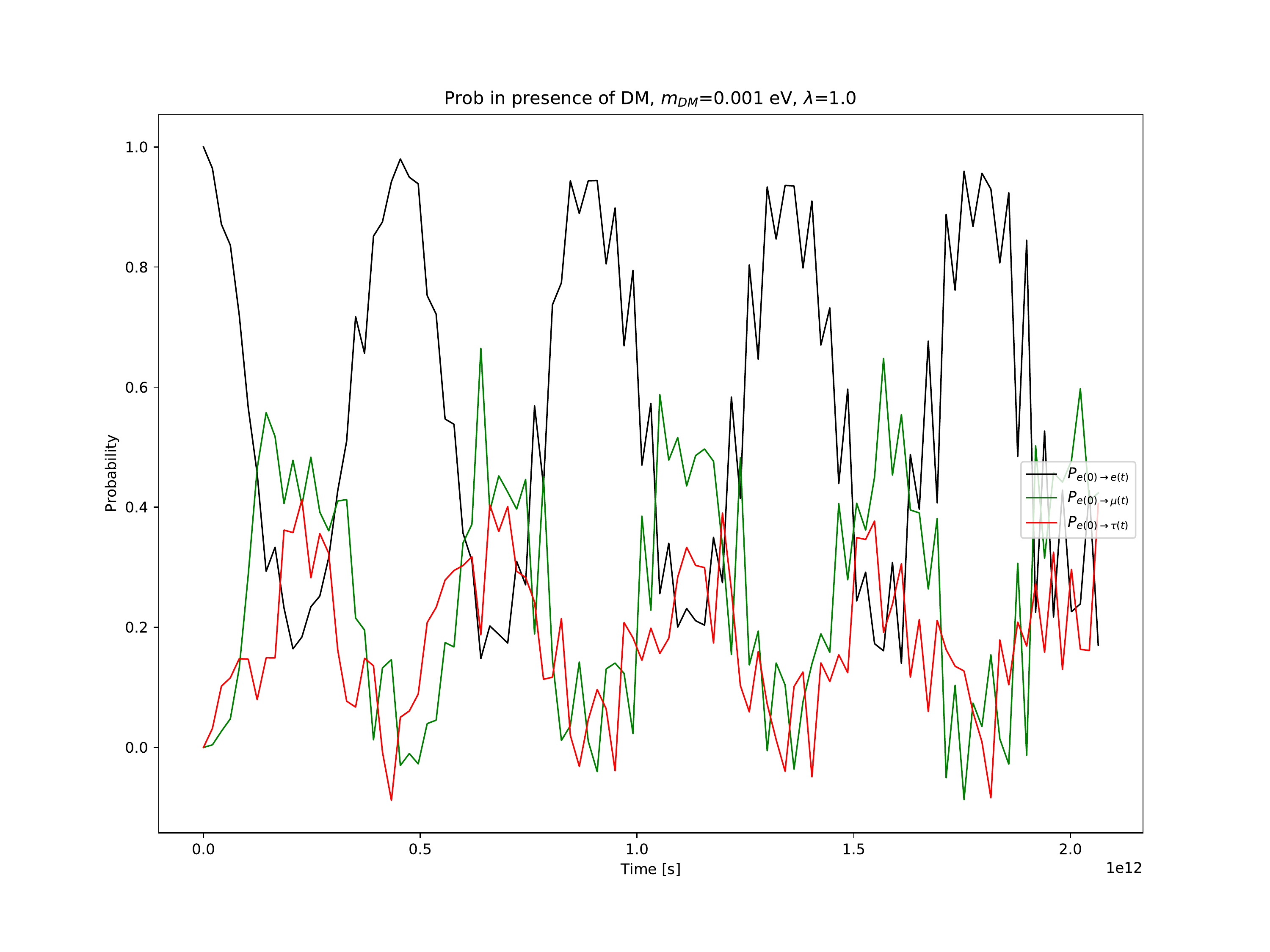}
\label{fig:subfigaa}}
\subfloat[][]{
\includegraphics[width=0.33\textwidth]{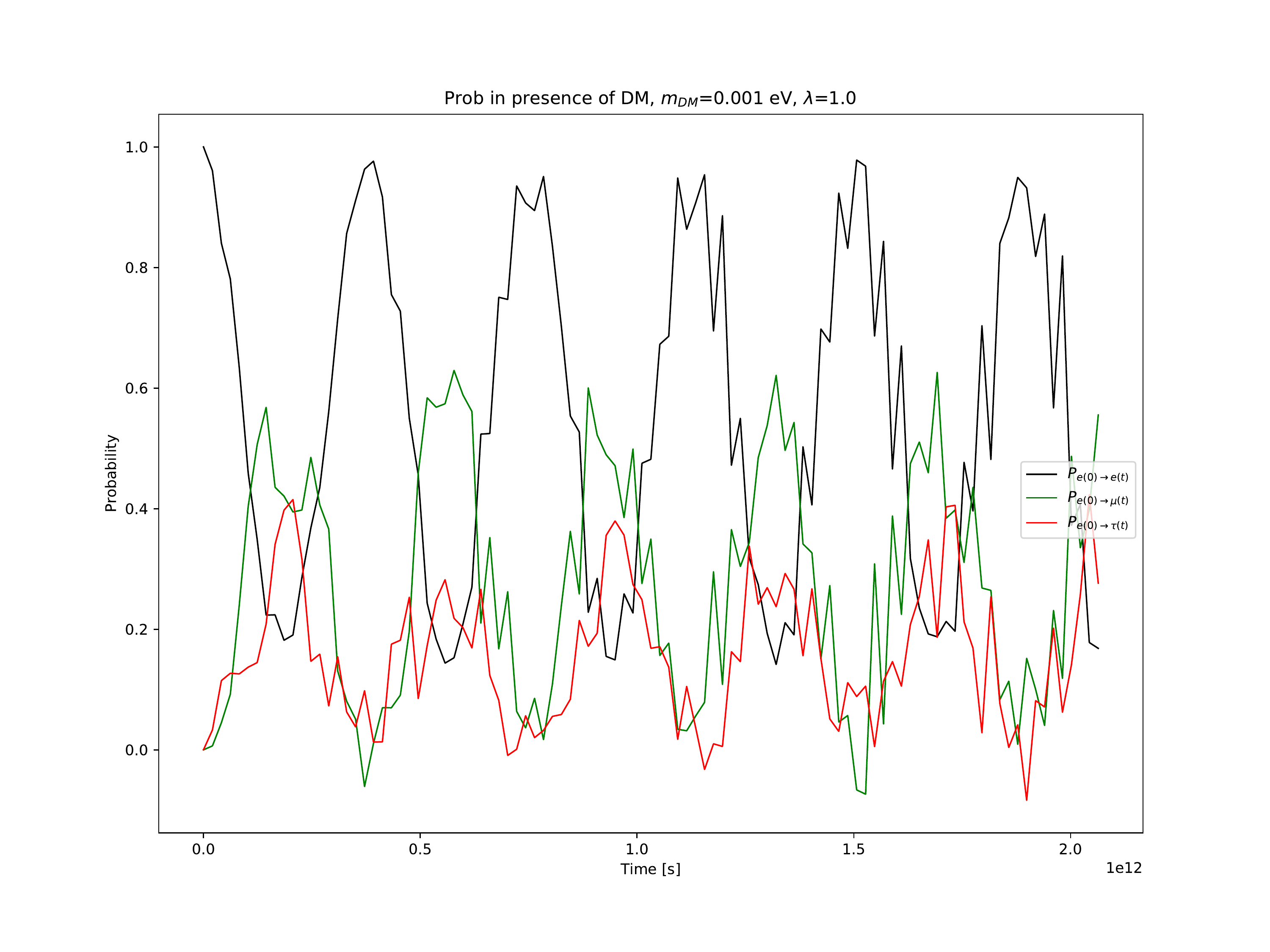}
\label{fig:subfiga}}
\subfloat[][]{
\includegraphics[width=0.33\textwidth]{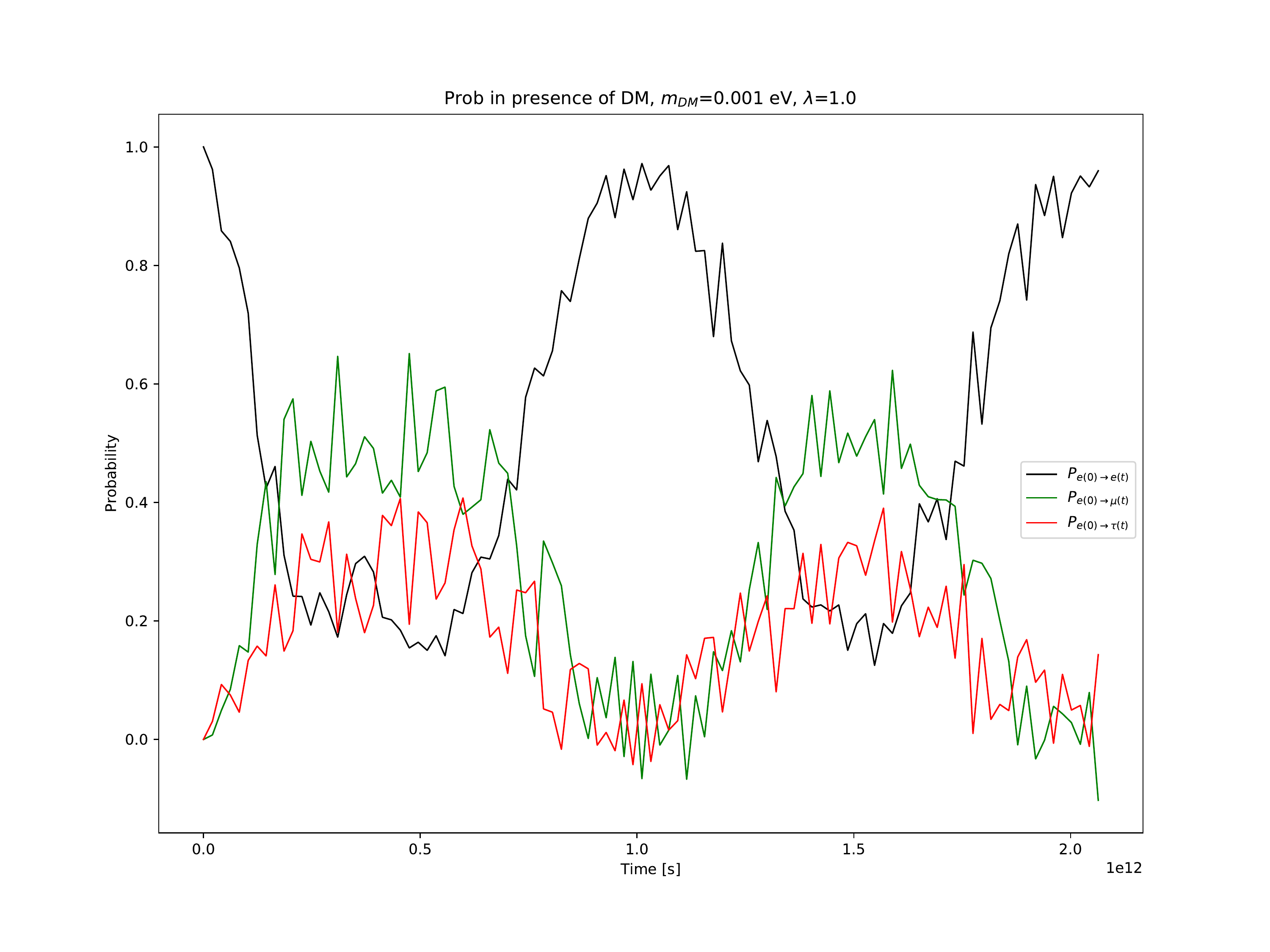}
\label{fig:subfigb}}
\qquad  
\subfloat[][]{
\includegraphics[width=0.33\textwidth]{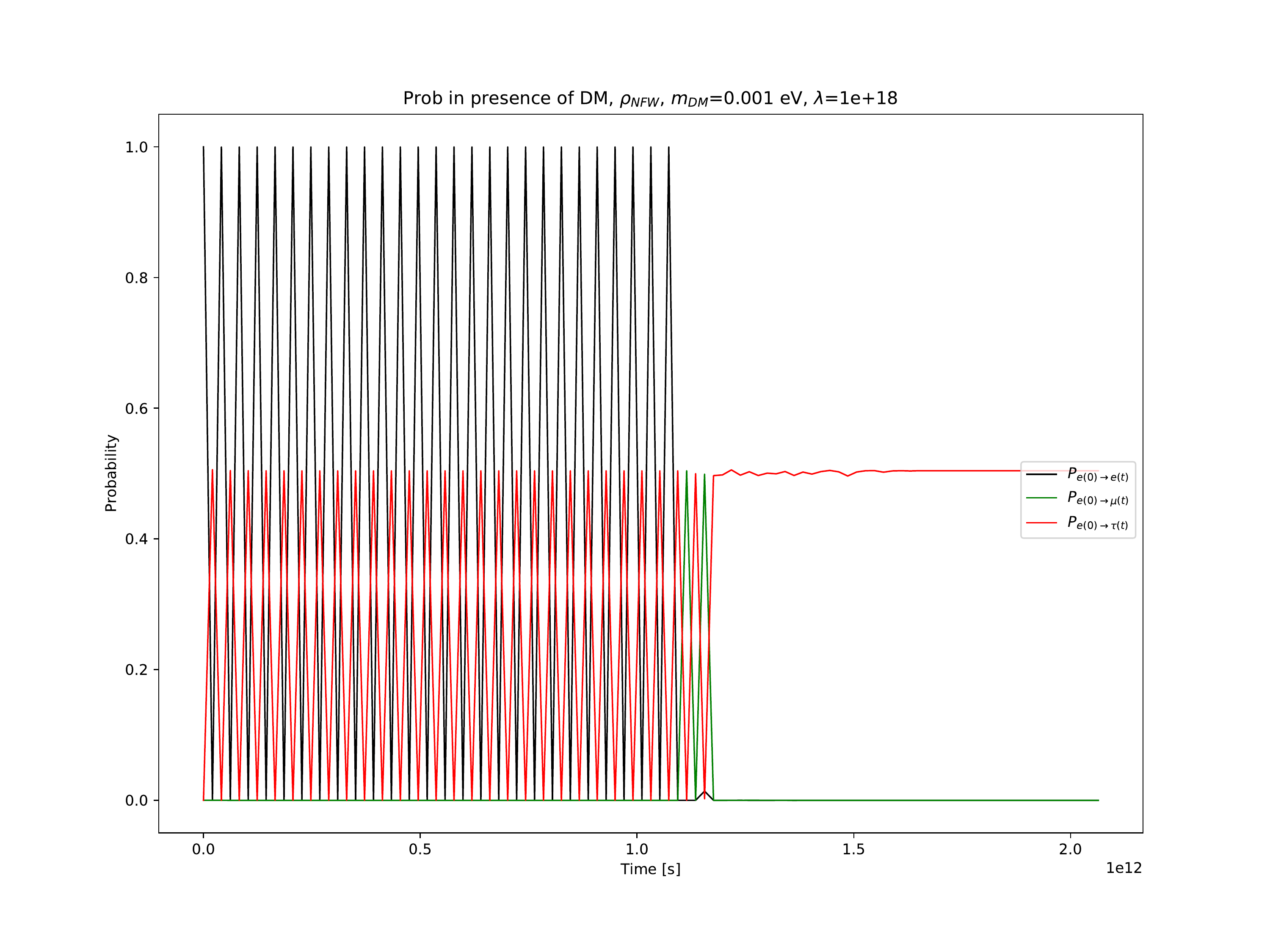}
\label{fig:subfigj}}
\subfloat[][]{
\includegraphics[width=0.33\textwidth]{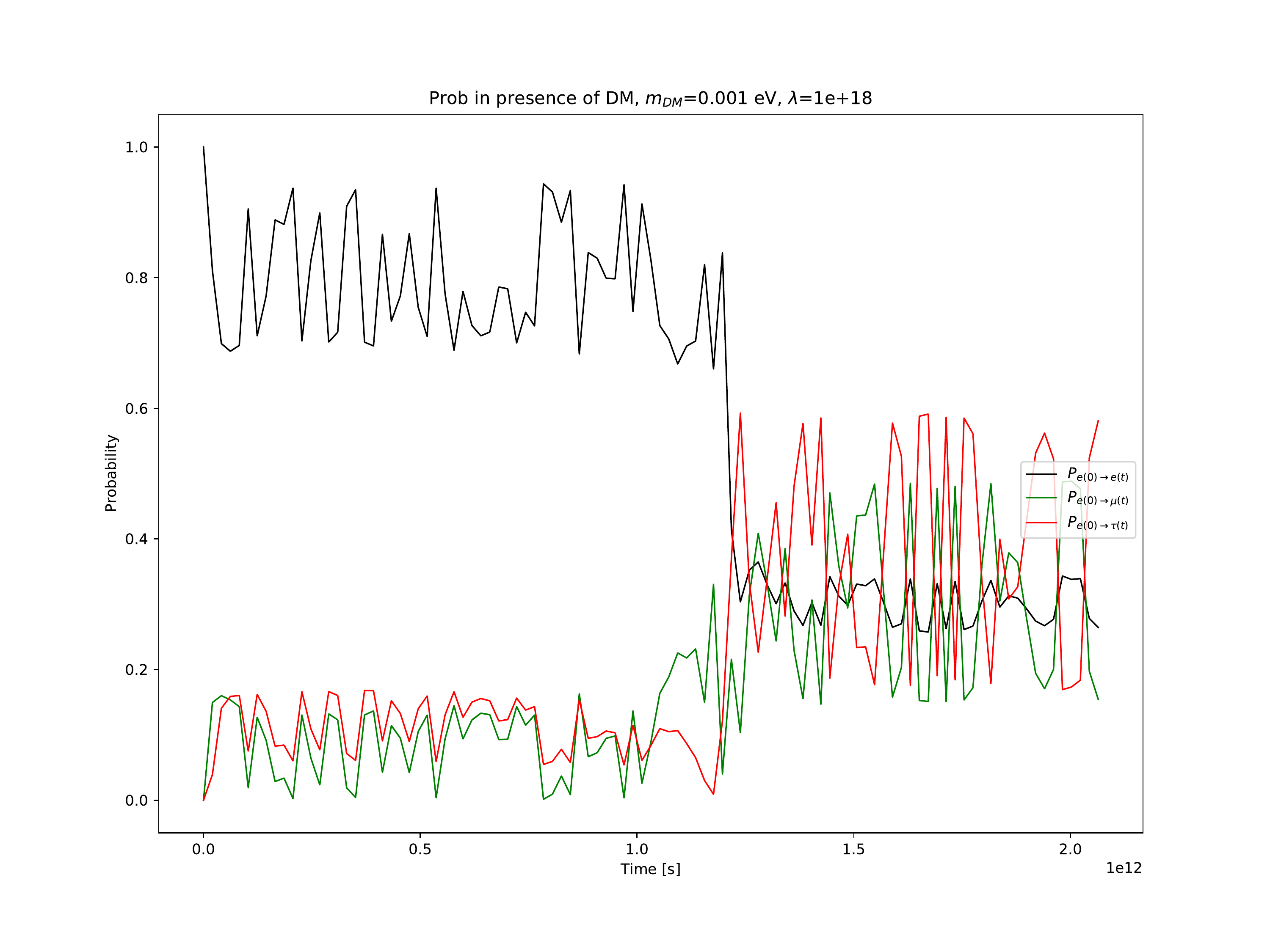}
\label{fig:subfigk}}
\subfloat[][]{
\includegraphics[width=0.33\textwidth]{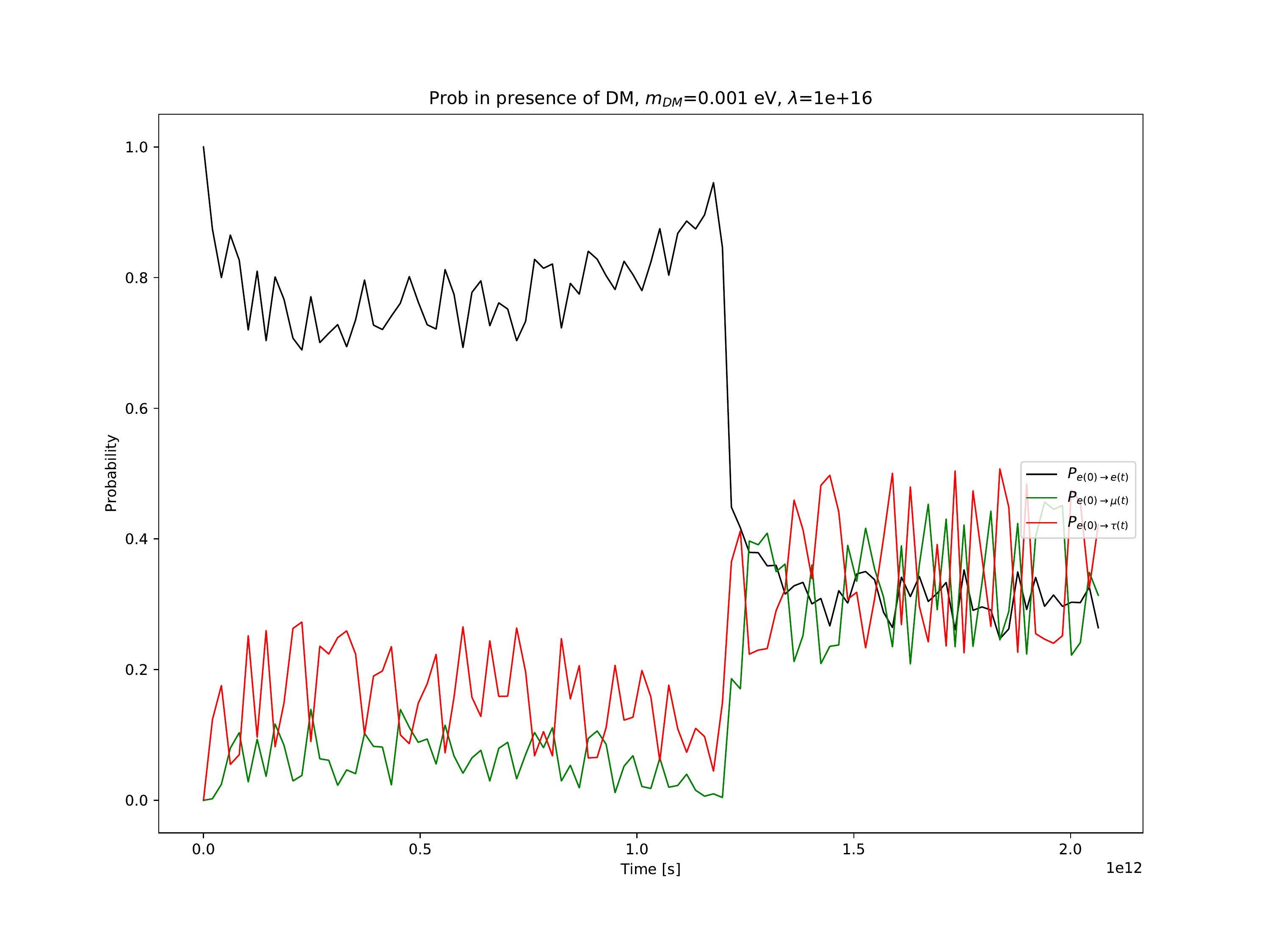}
\label{fig:subfigq}}
\qquad  
\subfloat[][]{
\includegraphics[width=0.33\textwidth]{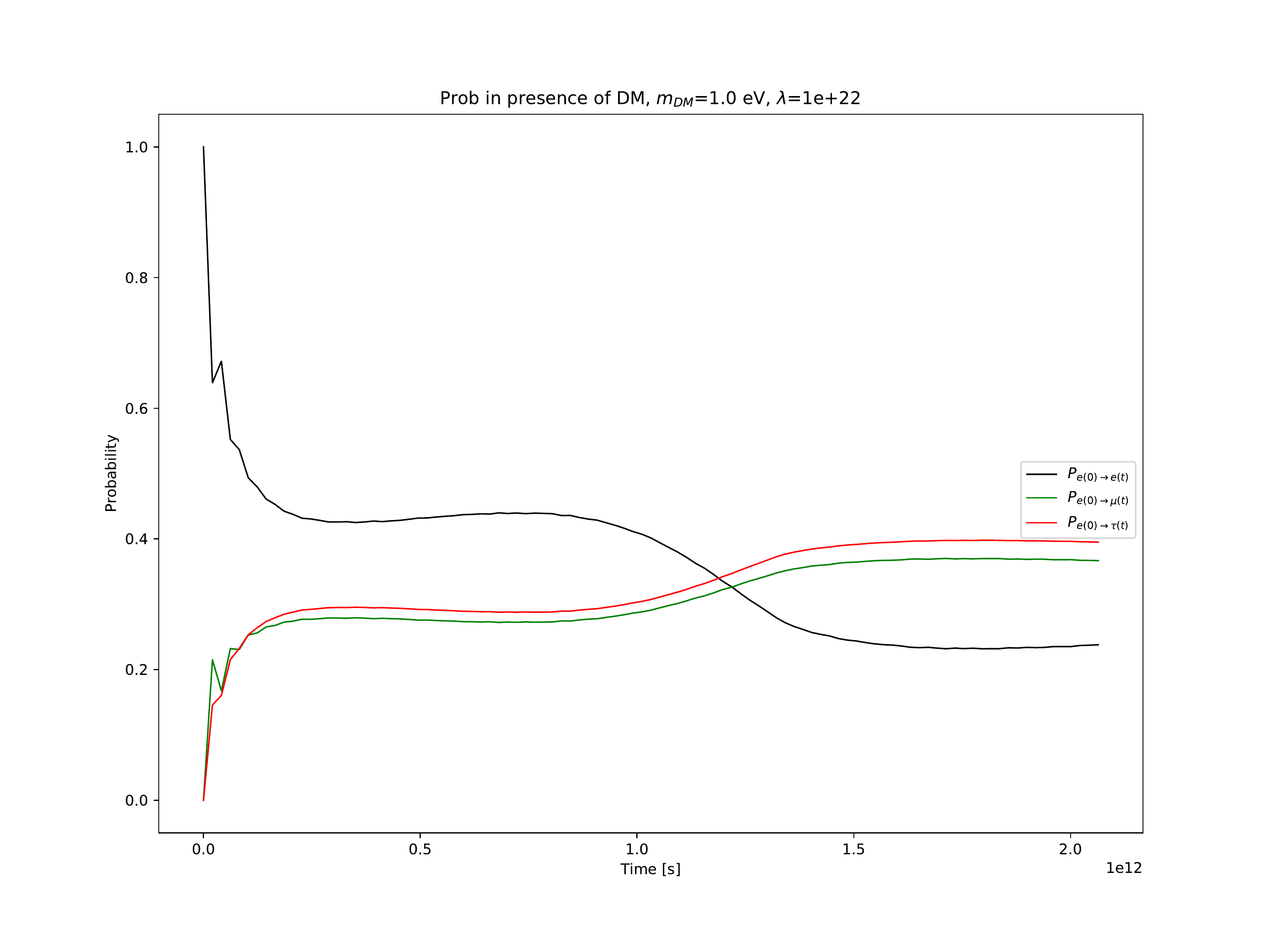}
\label{fig:subfigm}}
\subfloat[][]{
\includegraphics[width=0.33\textwidth]{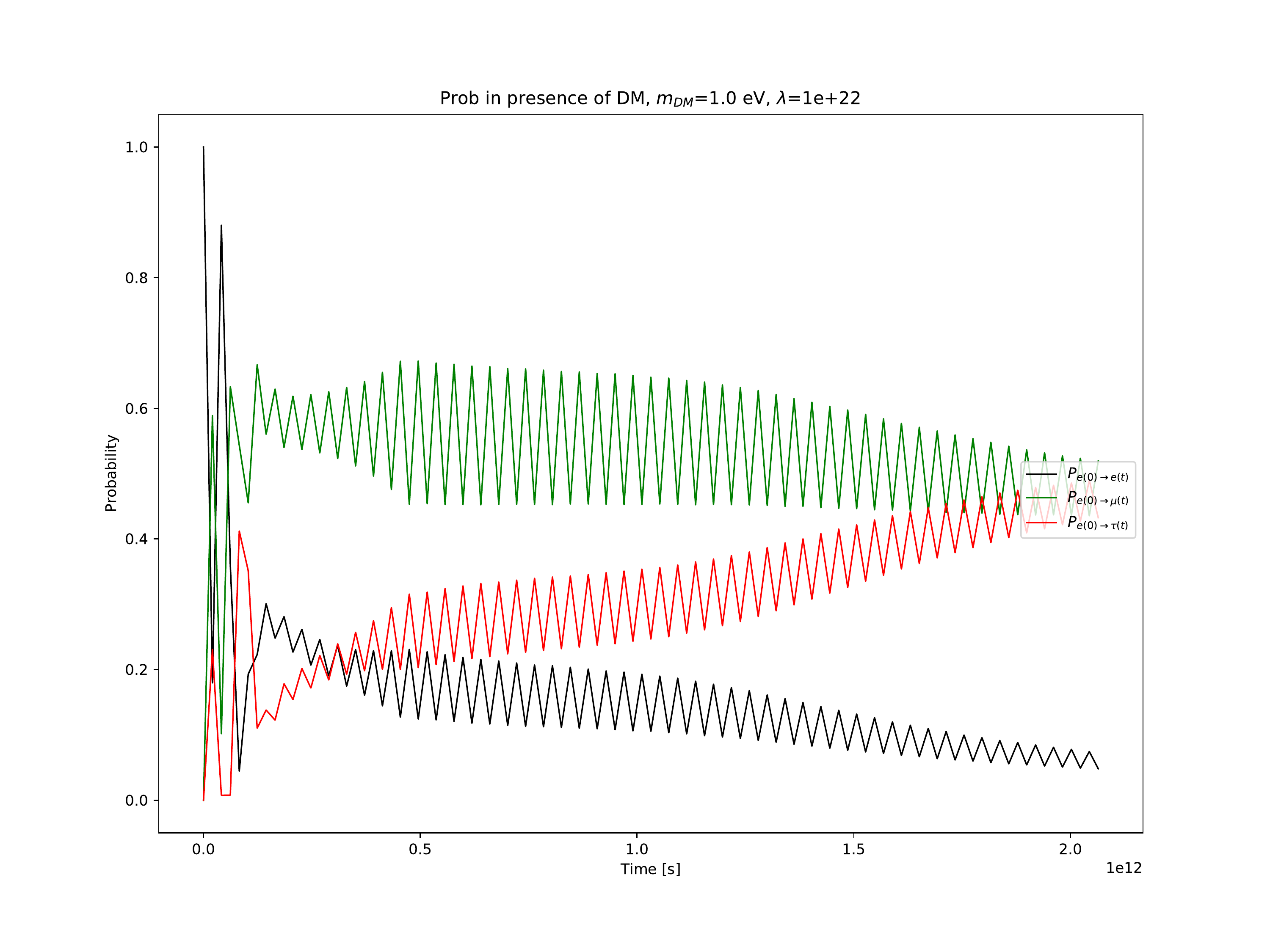}
\label{fig:subfigp}}
\subfloat[][]{
\includegraphics[width=0.33\textwidth]{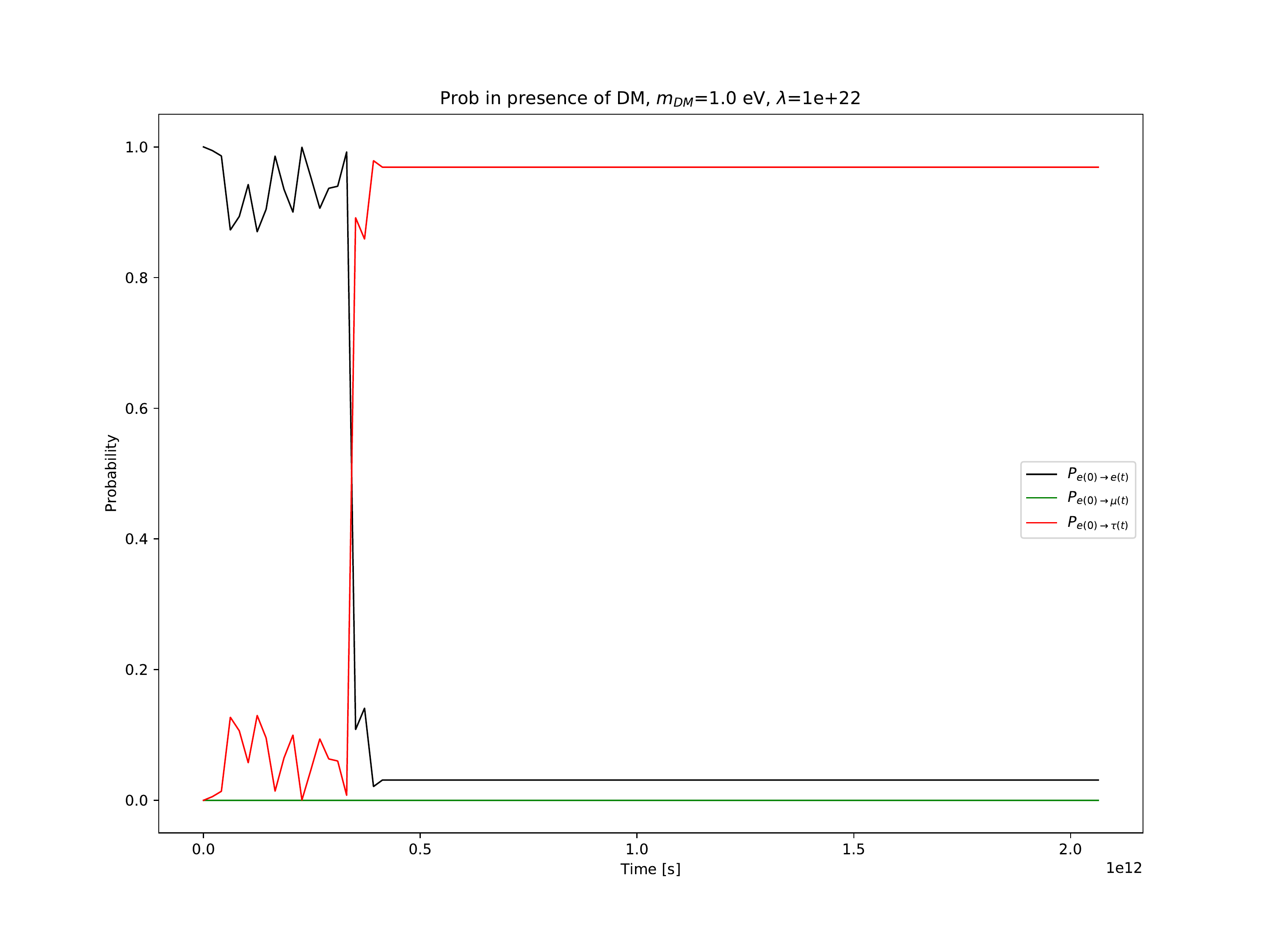}
\label{fig:subfigl}}
\qquad  
\caption{Survival and disappearance probabilities in presence of DM as a function of time, considering different values of the DM mass $m_{DM}$ and the dimensionless parameter $\lambda$, different DM density profiles and different textures for the matrix $\Lambda$.}
\label{fig:prob_V}
\end{figure*}

With $\Lambda_{(a)}$ we introduce a dependence of the mass eigenstates upon the DM properties (see Eq. \ref{H_m'}), opening the possibility for a MSW effect. $\Lambda_{(b)}$ may partially suppress components of the neutrino density matrix of a given flavor and $\Lambda_{(c)}$ activates all channels of the interaction. The results of the calculations done with these values of $\Lambda$ are shown in Figures \ref{fig:subfigaa}, \ref{fig:subfiga} and \ref{fig:subfigb}, respectively. For $\Lambda_{(a)}$ there are no noticeable effects on the oscillation pattern with respect to the case in vacuum. Instead, we notice an increase of the frequency of the oscillations in the case of $\Lambda_{(b)}$ and a decrease in the case of $\Lambda_{(c)}$.

Next, we keep the same value for $m_{DM}$, we set $\lambda=10^{18}$ and take the NFW-DM density distribution $\rho_{\rm{NFW}}(r)$ (see Eq. \ref{eq:NFW}). For this value of $\lambda$ the effect of the interaction is negligible at the source but it gains importance as the neutrino travels to the detector, reaching and even exceeding the order of magnitude of the initial mass eigenvalues. The results for this choice of parameters and for the texture of the matrix $\Lambda$ 
\newline
$\Lambda_{(d,e,f)}= 
\begin{bmatrix}
1 & 0 & 0\\
0 & 0 & -1\\
0 & 1 & 0\\
\end{bmatrix}$,
$\begin{bmatrix}
1 & 0 & 0\\
0 & 1 & 0\\
0 & 0 & -1\\
\end{bmatrix}$, and
$\begin{bmatrix}
1 & 0 & 0\\
0 & 0 & 1\\
0 & 1 & 0\\
\end{bmatrix}$
\newline
are shown in Figures \ref{fig:subfigj}, \ref{fig:subfigk} and \ref{fig:subfigq}, respectively. In these cases, the local dependence of the DM density profile, $\rho_{\rm{NFW}}$, induces huge effects when the neutrinos cross the GC at $r=0$ around the time $t \sim 1.18 \times 10^{12}$ s. The non-diagonal structure of the matrix $\Lambda$ for cases (d) and (f) produces a noticeable effect transforming the density matrix of the electron-neutrino from pure to mixed, thus resulting in decoherence. In case (e), as $\Lambda$ consists of diagonal elements, the oscillation pattern is dominated by the MSW effect, as was the case for the previous set of figures. 

Finally, we take $m_{DM}=1$ eV, $\lambda=10^{22}$, the isotropic density distribution $\rho_{\rm{iso}}(r)$ (see Eq. \ref{eq:iso}) and the following textures \newline $\Lambda_{(g,h,i)}= 
\begin{bmatrix}
0 & 0 & 0\\
0 & 0 & -1\\
0 & -1 & 0\\
\end{bmatrix}$,
$\begin{bmatrix}
1 & 0 & 0\\
0 & 0 & 1\\
0 & 0 & 0\\
\end{bmatrix}$, and
$\begin{bmatrix}
-1 & 0 & 0\\
0 & 0 & 0\\
0 & 0 & -1\\
\end{bmatrix}$.\\
The results for these values are shown in Figures \ref{fig:subfigm}, \ref{fig:subfigp} and \ref{fig:subfigl}, respectively. 
In case (g), as $V$ has only non-diagonal terms, decoherence dominates over oscillations. In case (h) there are mixed elements. This leads to a combined effect, in which oscillations are present with a very high frequency, but globally the three states tend to a pointer state. In case (i) the interaction is repulsive, leading to the suppression of the oscillations and the appearance of marked pointer states.  

\begin{figure*}
\centering
\subfloat[][]{
\includegraphics[width=0.33\textwidth]{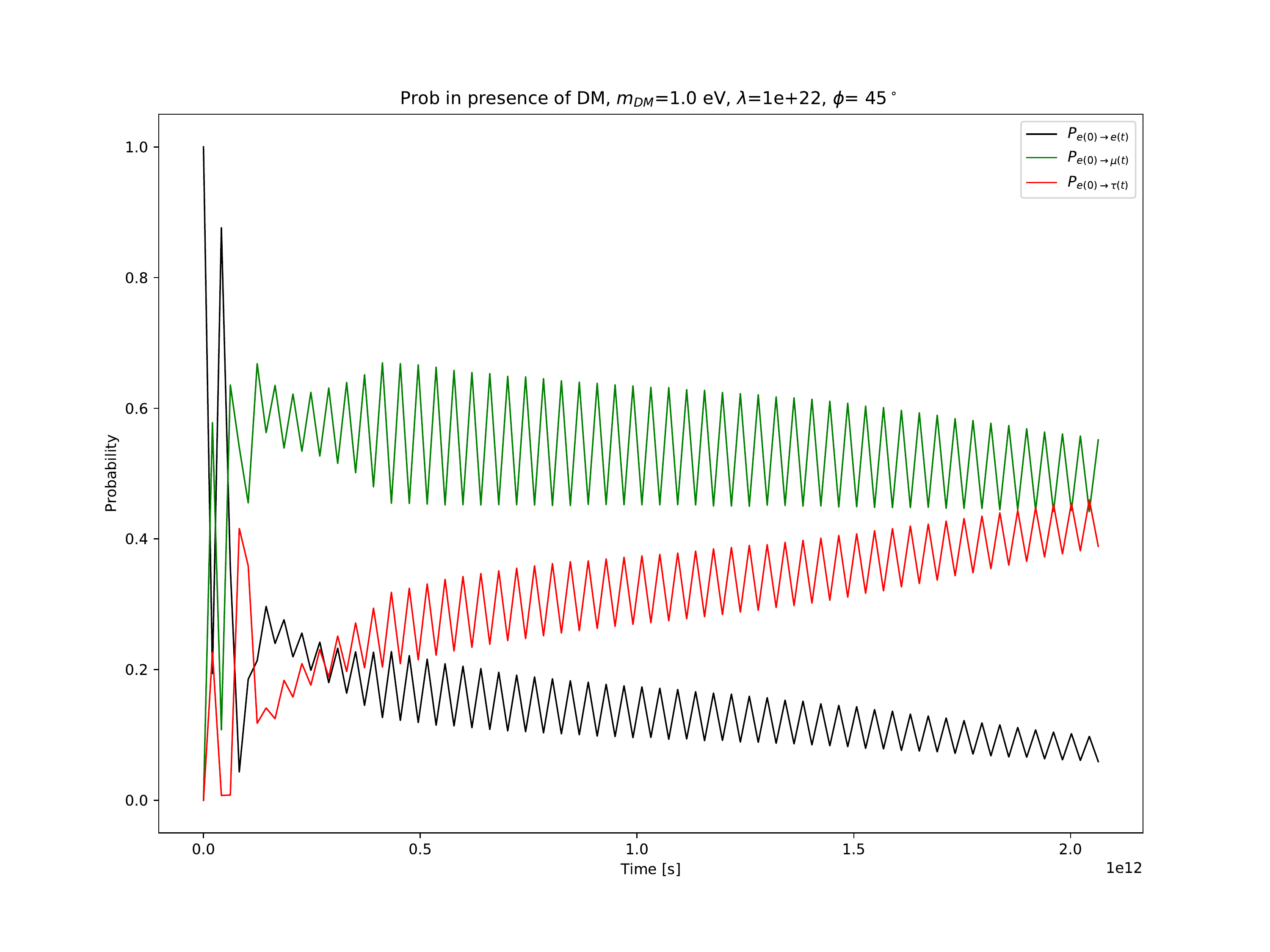}
\label{fig:phi45}}
\subfloat[][]{
\includegraphics[width=0.33\textwidth]{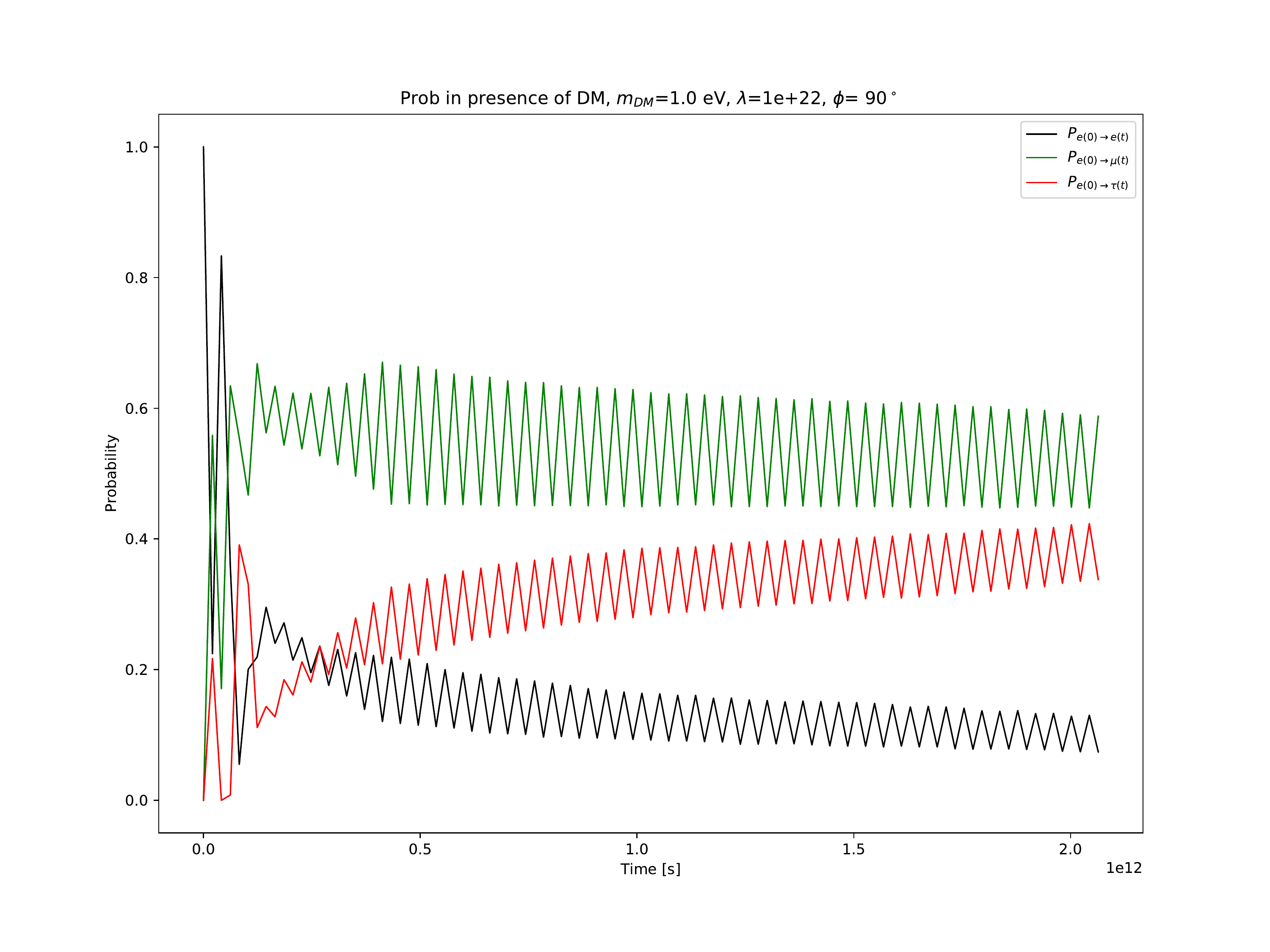}
\label{fig:phi90}}
\subfloat[][]{
\includegraphics[width=0.33\textwidth]{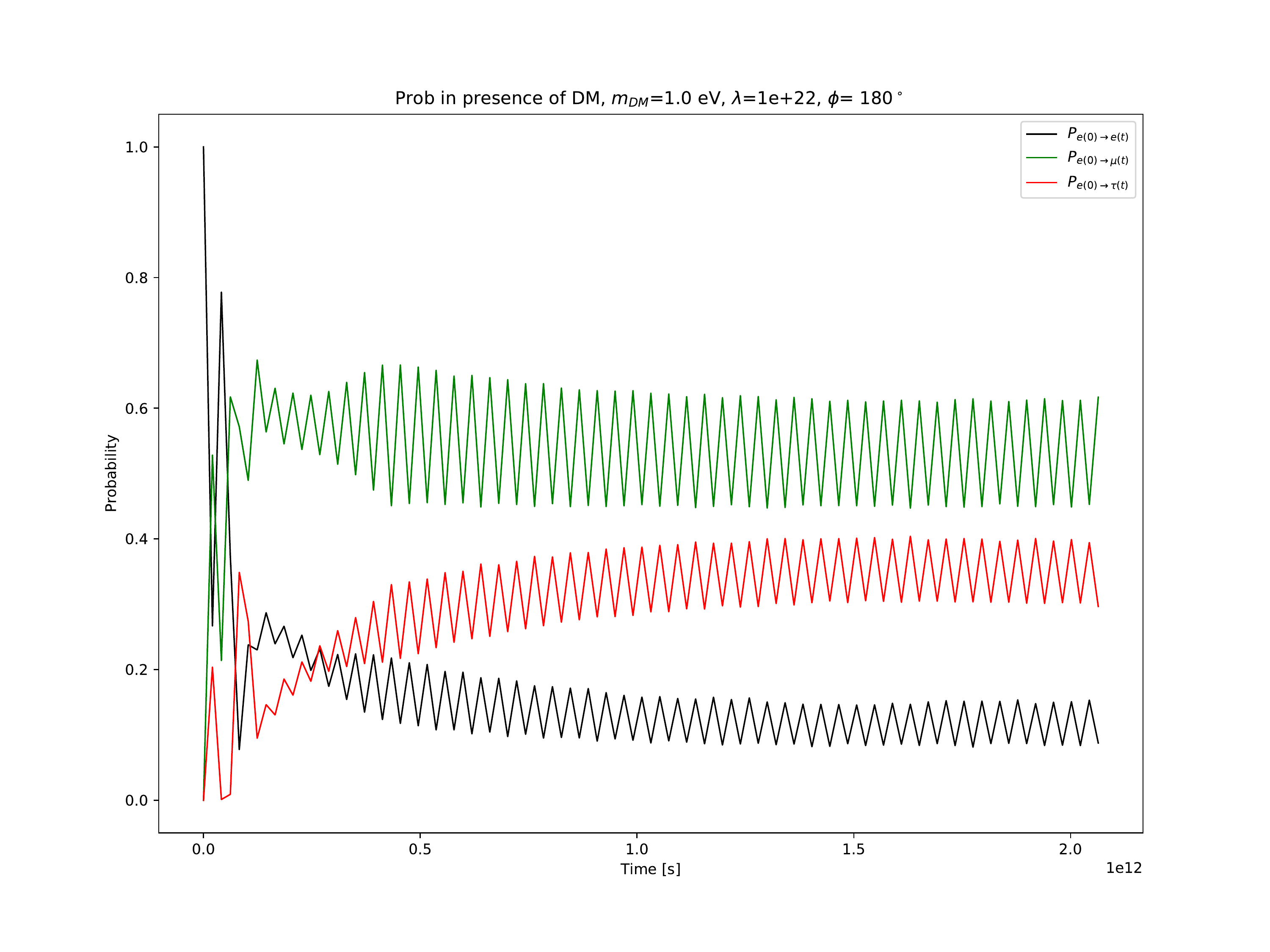}
\label{fig:phi180}}
\caption{Calculation of the survival and disappearance probabilities as a function of time for different values of the angle $\phi$: (a) $45^\circ$, (b) $90^\circ$ and (c) $180^\circ$. The other parameters are the ones of Figure \ref{fig:subfigp}.}
\label{fig:phi}
\end{figure*}

Concerning the dependence on the angle $\phi$, one can see from Eq. \ref{eq:r} that for values of $\phi$ larger than a few cents of a degree, the neutrino path does not cross the GC, which is the zone where the DM density distribution reaches its maximum value. For these trajectories the effect due to the presence of DM decreases.  To illustrate this dependence on $\phi$, we show in Figure \ref{fig:phi} results corresponding to $\phi=45$, $90$ and $180$ degrees.  

\section{Conclusions}\label{sec:conclusions}

In this work we have explored the effects of the interactions of neutrinos emitted from a distant source with a background of DM. For the DM distribution we have taken some of the most widely accepted profiles. We have varied all the parameters associated with the neutrino-DM interactions and found that the neutrino-flavor survival (disappearance) probabilities are sensitive to these parameters. Depending on the choice of the DM density, the probabilities may show  signals of the occurrence of the MSW effect and the neutrino-flavor states may also evolve to pointer states due to the onset of decoherence. 

In spite of the unknowns concerning DM properties, it seems reasonable to assess the role of extragalactic neutrinos as tracers of DM. As shown by our results, changes in the flavor composition of neutrinos emitted in distant sources can be attributed to the presence of DM, once the emission mechanism is fixed.

\acknowledgments

This work has been partially supported by the National Research Council of Argentina (CONICET) by the grant PIP 616, and by the Agencia Nacional de Promoci\'on Cient\'ifica y Tecnol\'ogica (ANPCYT) PICT 140492. A.V.P and O.C. are members of the Scientific Research career of the CONICET.

\end{document}